\documentclass[aip,jcp,reprint,amssymb,amsmath,superscriptaddress,groupedaddress,frontmatterverbose,]{revtex4-2}

\usepackage[title]{appendix}
\usepackage{graphicx, tikz, orcidlink}
\usepackage{epstopdf}
\usepackage{ascmac}
\usepackage{ulem}
\usepackage{cancel}
\usepackage{here}
\usepackage{physics}
\usepackage{bm}
\usepackage{comment}
\usepackage{booktabs}

\DeclareGraphicsExtensions{{.eps,.pdf,.png}}
\graphicspath{{./}}

\begin{document}

\title{HEOM-Based Numerical Framework for Quantum Simulation of Two-Dimensional Vibrational Spectra in Molecular Liquids (HEOM-2DVS)}
\date{Last updated: \today}

\author{Ryotaro Hoshino\orcidlink{0009-0004-4208-326X}}
\author{Yoshitaka Tanimura\orcidlink{0000-0002-7913-054X}}
\email[Author to whom correspondence should be addressed: ]{tanimura.yoshitaka.5w@kyoto-u.jp}
\affiliation{Department of Chemistry, Graduate School of Science,
Kyoto University, Kyoto 606-8502, Japan}

\begin{abstract}
The multi-mode anharmonic Brownian motion model provides a universal framework for simulating molecular vibrations in condensed phases. When vibrational energy surpasses thermal excitation, quantum effects become significant, necessitating a rigorous treatment of system–bath entanglement. The hierarchical equations of motion (HEOM) provide a powerful methodology for simulating such open quantum systems. In this context, two‑dimensional vibrational spectroscopy (2DVS) constitutes a powerful probe for elucidating the complex dynamics of molecular processes, both experimentally and theoretically. This work introduces a computational implementation, HEOM‑2DVS, for treating non‑Markovian open quantum dynamics that encompass energy relaxation, dephasing, thermal excitation, and related processes arising from non‑perturbative and nonlinear interactions between selected vibrational modes and their thermal environments. To validate the theoretical framework, we computed 2D correlation infrared spectra for three coupled intramolecular vibrational modes of water. The HEOM-2DVS program developed for both CPU and graphics processing unit (GPU) is provided as supplementary material.
\end{abstract}

\maketitle

\section{INTRODUCTION}

The vibrational dynamics of molecules in condensed phases have increasingly been recognized as crucial factors shaping chemical reactivity. In particular, intramolecular motions in solution---most prominently the OH stretching vibration of water---have garnered significant interest as active contributors to reactivity.\cite{covington_physical_chemistry_1978,mabesoone_solute_2022}
Rather than serving as a passive thermal background, molecular environments exhibit ultrafast phenomena, including energy and phase relaxation. In hydrogen-bonding solvents, these dynamics may further involve hydrogen-bond rearrangement and proton migration, all of which complicate its analysis.\cite{Ohmine_ChemRev93,OCSACR1999,Nibbering2004UltrafastVD,bagchi_2013} 

To elucidate these dynamics, femtosecond-resolved measurements are pivotal, as they coincide with the intrinsic timescales of fundamental chemical processes, encompassing bond rearrangements and chemical reactions.
Two-dimensional vibrational spectroscopy (2DVS) has established itself as a powerful tool for resolving vibrational mode correlations, coherence lifetimes, and pathways of energy flow with exceptional spectral precision.\cite{mukamel1999principles,HammLimRobin1998,HammLimmDeGradoHochstrasser2000,HammHochstrasser2001_2DIRReview,Demirdoven-Khalil-Tokmakoff:PRL2003,Cho2009,Hamm2011ConceptsAM,HammPerspH2O2017} 
Note that because our model-based framework provides a unified treatment of infrared (IR) and Raman spectroscopies,\cite{IIT15JCP,IT16JCP,HT25JCP1,HT25JCP2,TT23JCP1,TT23JCP2}  we adopt the more general term 2DVS, even though the present study focuses specifically on 2D IR calculations.
Its acute sensitivity to anharmonicity, mode coupling, and vibrational coherence allows rigorous quantification of the relaxation–dephasing mechanisms that govern spectral broadening.\cite{SkinnerFayerJCP2004,ElsaesserH2O,TokmakoffNat2013,HammTHz2012,Blake20202DThzRaman,Begusic2023} Since these nonlinear spectral signatures are intimately tied to quantum dissipative dynamics---phenomena beyond the current reach of molecular dynamics (MD) simulations---robust theoretical modeling remains indispensable for their interpretation and full exploitation.\cite{TI09ACR}

For decades, MD simulations have served as a principal framework for investigating the dynamical properties of solutions.\cite{Shinji2DRaman2006,HT06JCP,LHDHT08JCP,HT08JCP,YagasakiSaitoJCP20082DIR,Saito1995water3modes,Saito1997water3modes,YagasakiSaitoJCP2011Relax,Yagasaki_ARPC64,Wei2015Nagata2DRamanTHz,IHT14JCP,JIT16CP,IHT16JPCL,ImotXanteasSaitoJCP2013H2O,Imotobend-lib2015} Classical MD, however, is intrinsically incapable of incorporating essential quantum mechanical phenomena---zero-point energy, tunneling, and quantum thermal fluctuations---that are indispensable for a faithful description of vibrational dephasing and couplings among intermolecular modes. Quantum MD methodologies, such as path-integral Centroid MD (\texttt{PI-CMD}), have been advanced to address these deficiencies; yet their application to 2DVS remains computationally formidable.\cite{Babin2013_MBpol_I,Babin2014_MBpol_II,Medders2014_MBpol_III}
To confront these challenges, MD-based modeling frameworks have been advanced, incorporating stochastic dynamics\cite{SkinnerStochs2003,Skinner2004HOD,Skinner2005HOD,Skiner2005nonCondon} and excitonic wavefunction approaches.\cite{Mukamel2009,Mukamel2009Spectron,Jansen2014NISE} Machine-learning (ML) methodologies leveraging MD trajectories have likewise been developed.\cite{UT20JCTC,PJT25JCP1,PUT26JCP1} 

For accurate modeling of vibrational dephasing and relaxation, nonlinear, non-perturbative, and non-Markovian system-bath (S-B) interactions must be incorporated, since the vibrational echo signal originates from S-B entanglement.\cite{T06JPSJ,T20JCP} Our group has performed multidimensional spectral analyses using the multimode anharmonic Brownian (MAB) model,\cite{TI09ACR,ST11JPCA} and developed hierarchical Fokker–Planck equations (\texttt{HFPE}) in both classical (\texttt{CHFPE})\cite{IIT15JCP,IT16JCP,HT25JCP1,HT25JCP2} and  quantum (\texttt{QHFPE})\cite{TW91PRA,T15JCP} form, and discretized hierarchical equations of motion in mixed Liouville–Wigner space (\texttt{DHEOM-MLWS}).\cite{TT23JCP1,TT23JCP2}  These enable numerically precise simulations of nonlinear spectra in complex systems. By calibrating \texttt{CHFPE} to reproduce classical MD benchmarks\cite{IT16JCP,PUT26JCP1} and applying \texttt{QHFPE} for quantum-level insights,\cite{TT23JCP1,TT23JCP2} the quantum nature of vibrational dynamics is revealed. In contrast, classical simulations remain suitable for 2D intermolecular spectroscopies---such as 2D Raman\cite{Shinji2DRaman2006,HT06JCP,LHDHT08JCP} and 2D THz-Raman spectroscopy\cite{IHT14JCP,IIT15JCP,JIT16CP,IHT16JPCL,HT25JCP2}---where thermal excitation suppresses quantum coherence.

For intramolecular modes exhibiting significant quantum effects, computational approaches to 2DVS have thus far been developed within the \texttt{DHEOM-MLWS} framework for two-mode MAB models, typically involving stretching and bending vibrations.\cite{TT23JCP1,TT23JCP2} While such models capture mode–mode coupling, a three-mode formulation is required to describe energy transfer pathways and coherence dynamics. Given the experimental precision of 2DVS in resolving these processes, extending to three-mode models is indispensable. We previously carried out classical simulations of an MAB system, incorporating the symmetric, asymmetric, and bending vibrational modes\cite{HT25JCP1,HT25JCP2}---but these results highlight the limitations of classical treatments in fully accounting for ultrafast coherence-driven relaxation.\cite{ST11JPCA} Therefore, in this work, we present an HEOM‑based computational framework for simulating 2D correlation IR spectra\cite{Demirdoven-Khalil-Tokmakoff:PRL2003,Khalil-Demirdoven-Tokmakoff:JPCA2003,IT06JCP,IT07JPCA,2DCrrJonas2001,2DCrrGe2002,2DCrrTokmakoff2003,T12JCP} extending previous approaches to treat three interacting intramolecular modes within an open quantum dynamics setting.\cite{TT23JCP1,TT23JCP2,HT25JCP1,HT25JCP2} The resulting implementation, \texttt{HEOM‑2DVS}, enables non‑Markovian simulations that capture energy relaxation, dephasing, thermal excitation, and related effects arising from non‑perturbative and nonlinear mode–bath interactions.

Unlike existing 2D simulation packages such as \texttt{SPECTRON},\cite{Mukamel2009Spectron} \texttt{NISE},\cite{Jansen2014NISE} and \texttt{g$\_$spec},\cite{Tokmakoff2000gSpec} the \texttt{HEOM‑2DVS} approach incorporates anharmonic mode–mode coupling together with a fully non‑perturbative, non‑Markovian treatment of system–bath interactions. This capability is essential for capturing fluctuation–dissipation effects at finite temperature and for accurately describing systems such as liquid water. Although several HEOM implementations have been developed recently,\cite{Shi2009HEOM,SHI10.1063/1.5026753,IKEDASCHOLES2020,IKEDANAKAYAMA2022,Huang2023,Ke2023TreeTensorHEOM,Borrelli2021,Takahashi2024,Ignacio2024,Ignacio2025,LinjunWan2025} they still face limitations in treating complex interactions—particularly vibrational dephasing—which is the central focus of this study. In this paper, we present a computational software developed for simulating 2D IR spectra based on HEOM framework, employing an efficient numerical algorithm accelerated by a graphics processing unit (GPU).

 This paper is organized as follows. Section~\ref{sec:theory} introduces the MAB model and the HEOM for intramolecular vibrational modes. Section~\ref{sec:HEOMcode} briefly describes the structure of our codes, and Section~\ref{sec:Demo} demonstrates their capability through simulations of linear abosorption spectra and 2D correlation IR spectra. Concluding remarks are provided in Section~\ref{sec:conclude}.

\section{MAB model and HEOM}
\label{sec:theory}

\subsection{MAB model}
\label{sub:MMBO}
We consider a model consisting of three primary intramolecular modes.  
These modes are described by vibrational coordinates $\bm{q}=(q_1, q_{2}, q_3)$. Each mode is independently coupled to the other optically inactive modes, which constitute a bath system represented by an ensemble of harmonic oscillators. The total Hamiltonian can then be expressed as\cite{UT20JCTC,PJT25JCP1,PUT26JCP1,IIT15JCP,IT16JCP,TT23JCP1,TT23JCP2,HT25JCP1,HT25JCP2} 
\begin{align}
\hat{H}_{tot} &=  \sum_{s} \left( \hat{H}_{A}^{(s)}
+\hat{H}_{I}^{(s)} +\hat{H}_{B}^{(s)} +\hat{H}_{C}^{(s)}  \right) \nonumber \\
&+  \sum_{s<s'} \hat{U}_{ss'}\qty(\hat{q}_s, \hat{q}_{s'}),
\label{sec:Total Hamiltonian}
\end{align}
where
\begin{align}
\hat{H}_{A}^{(s)}= \frac{\hat{p}_s^{2}}{2m_s} +\hat U_s(\hat{q}_s)
\label{sec:System Hamiltonian}
\end{align}
is the Hamiltonian for the $s$th mode, with mass $m_s$, coordinate ${\hat{q}_s}$, and momentum ${\hat p_s}$; and
\begin{align}
\hat U_s(\hat{q}_s)= \frac{1}{2} m_s \omega_s^2 \hat{q}_s^2 +\frac{1}{3!}g_{s^3}q_{s}^3
\label{sec: Potenentials}
\end{align}
is the anharmonic potential for the $s$th mode, described by the frequency  $\omega_s$ and cubic anharmonicity $g_{s^3}$. 
The anharmonic coupling between the $s$th and $s'$th modes is given by 
\begin{align}
\hat{U}_{ss'}(\hat{q}_s, \hat{q}_{s'}) = g_{s{s'}}\hat{q}_s\hat{q}_{s'} + \frac{1}{6}  \qty(g_{s^2s'}\hat{q}_s^2 \hat{q}_{s'} + g_{s{s'}^2} \hat{q}_s \hat{q}_{s'}^2 ),
\label{sec: Potential ss'}
\end{align}
where $g_{s{s'}}$ represents the second-order harmonicity, and $g_{s^2s'}$ and $g_{s{s'}^2}$ represent the third-order anharmonicity.  
In the third-order response function considered below, the contributions from even-order anharmonicity vanish.\cite{OT97JCP1} Therefore, here we retain only the third-order anharmonic terms.

The bath Hamiltonian for the $s$th mode is expressed as\cite{CALDEIRA1983587,T06JPSJ,TW91PRA,TM93JCP,T98CP}
\begin{align}
\hat{H}_{B}^{(s)}= \sum_{j_s}\qty(\frac{\hat{p}_{j_s}^{2}}{2m_{j_s}}+\frac{m_{j_s}\omega_{j_s}^{2} \hat{x}_{j_s}^2}{2} ),
\label{sec: Bath Hamiltonian}
\end{align}
where the momentum, coordinate, mass, and
frequency of the $j_s$th bath oscillator are given by ${p}_{j_s}$, ${x}_{j_{s}}$, $m_{j_{s}}$ and
$\omega _{{j_s}}$, respectively.  
The counter term, which maintains the translational symmetry of the system in the case $\hat U_s(\hat{q}_s)=\hat{U}_{ss'} (\hat{q}_s, \hat{q}_{s'})$=0 is defined as\cite{TW91PRA,OT97PRE}
\begin{align}
\hat{H}_{C}^{(s)}=  \Lambda^{(s)}  \hat{V}_s^2(\hat{ q}_s)
\label{sec: counter Hamiltonian}
\end{align}
with the factor $\Lambda^{(s)} \equiv \sum_{j_s} {\alpha_{j_s}^2 }/{2m_{j_s} \omega _{j_s}^2 }$.  The S-B interaction is expressed as
\begin{align}
  {H}^{(s)}_{\mathrm{I}}&=- V_{s}(\hat {q_s})\sum _{j_s}\alpha _{j_s}{\hat x}_{j_s},
  \label{eq:h_int}
\end{align}
where $V_{s}({q_s})\equiv V^{(s)}_{\mathrm{LL}}{q_s}+V^{(s)}_{\mathrm{SL}}{q_s}^{2}/2$ with the linear-linear (LL)\cite{CALDEIRA1983587,TW91PRA,TM93JCP,T98CP}
and square-linear (SL) S-B interactions.\cite{TS20JPSJ,KT04JCP,OT97PRE} 
The coupling strengths are expressed by $V^{(s)}_{\mathrm{LL}}$, $V^{(s)}_{\mathrm{SL}}$, and $\alpha_{j_s}$. 
For a vibrational mode with weak anharmonicity, the LL interaction leads to energy relaxation, whereas the SL interaction results in vibrational dephasing.\cite{T06JPSJ,TS20JPSJ}

We consider optical measurements where the molecular system interacts with a laser field $E(t)$, while the effects of laser polarization are not included here.
The nonlinear elements of dipole are essential to 2D spectroscopy.  Here we assume\cite{IT16JCP,TT23JCP1,TT23JCP2,HT25JCP1,HT25JCP2}
\begin{eqnarray}
\hat {\mu} = \sum_s \mu^s \hat q_s + \frac{1}{2!}\sum_{s,s'}\mu^{ss'}\hat q_s \hat q_{s'},
 \label{NLdip}
\end{eqnarray}
where $\mu_s$ and $\mu_{ss'}$ are the linear and nonlinear elements of the dipole moment. For IR spectroscopies, the laser interaction is then expressed as $H_{\rm IR}(t)=- E(t){\mu}({\boldsymbol q})$. 

The system Hamiltonian can always be expressed in matrix form using the energy eigenstates of $\hat{H}_{A}^{(s)}$, denoted as
$\left| n_s \right\rangle$ with eigenenergy $ \hbar \omega _{n}^s= \left\langle {n_s } \right|  \hat{H}_{A}^{(s)} \left| {n_s' }\right\rangle$. Then for $\hat H_{S} \equiv \sum_{s} \hat{H}_{A}^{(s)} +\sum_{s<s'} \hat{U}_{ss'}\qty(\hat{q}_s, \hat{q}_{s'})$ we have
\begin{eqnarray}
\hat H_S &=& \hbar  \sum\limits_{s}  \sum\limits_{n} \omega _{n}^s  \left| {n_s } \right\rangle \left\langle {n_s } \right| 
\nonumber \\
&+& \hbar  \sum_{s<s'} \sum\limits_{n \ne {n'} } \sum\limits_{m_s \ne m' } \Delta _{n n'\, mm'}^{ss'} 
\left| {m}_{s'} \right\rangle \left| {n_{s} } \right\rangle  \left\langle {n_{s'}'} \right|  \left\langle {m'_{s'}} \right|, \nonumber \\
\end{eqnarray}
where $\hbar \Delta _{nn'\,mm'}^{ss'} =  \left\langle {n_{s} } \right|  \left\langle {m_{s'} } \right| \hat{U}_{ss'}(\hat{q}_s, \hat{q}_{s'}) 
\left| {n_{s}' } \right\rangle \left| {m'_{s'}} \right\rangle
$.

The dipole moment is now expressed as
\begin{equation}
  \hat{\mu} =  \sum_{s} \sum\limits_{n \ge {n'} } \mu_{n n' }^s  \left| {n_{s} } 
\right\rangle \left\langle n'_{s} \right| 
  + \sum_{s<s'} \sum\limits_{n_s \ne n' }  \mu _{n,n'}^{s,s'}  \left| {n}_{s} \right\rangle  \left\langle {n'_{s'} } \right| ,\nonumber \\
  \label{eq:mu}
\end{equation}
where $\mu_{n n' }^s= \left\langle {n_{s} } \right|  \mu^s \hat q_s  \left| {n' _{s}}\right\rangle$ and $ \mu _{n,n'}^{s,s'} =\mu^{ss'} \left\langle {n_{s} } \right|  \hat q_s \left| {n}_{s} \right\rangle  \left\langle {n'_{s'} } \right|  \hat q_{s'} \left| {n'_{s'} }\right\rangle/2$.

The total Hamiltonian is then given by
\begin{align}
\label{eq:total_hamiltonian}
\hat H_{tot} = & \hat H_\mathrm{S}' - \sum_{s} \sum_{j_s} \alpha_{j_s} \hat V_s \hat x_{j_s} \nonumber \\
 & +\sum_{s} \sum_{j_s} \left[ \frac{\hat{p}_{j_s}^{2}}{2m_{j_s}}+\frac{m_{j_s}\omega_{j_s}^{2}}{2} \hat{x}_{j_s}^2 \right],
\end{align}
where $\hat H_{S}' \equiv \hat H_{S} + \hat H_C$ and
\begin{align}
\label{eq:counter1}
\hat H_C= \hbar \sum_{s}  \sum\limits_{n \ge {n'} } \delta_{n {n' }}^{s} \left| {n_s } \right\rangle \left\langle {n_s' } \right|
\end{align}
with
\begin{align}
\label{eq:counter2}
\hbar  \delta_{n {n' }}^{s} = \Lambda^{(s)} \left\langle {n_s } \right| V_{s}^2 ({{\hat q}_s})  \left| {n_s' } \right\rangle.
\end{align}

The system part of the S-B interaction is expressed as 
\begin{equation}
\hat V_s =  \sum\limits_{n \ge {n'} }  V_{n n' }^s \left| {n_s } \right\rangle  \left\langle {n_s' } \right|,
\label{eq:sys_interaction}
\end{equation}
where $V_{n n' }^s \equiv \left\langle {n_s } \right| V_{s}({{\hat q}_s})  \left| {n_s' } \right\rangle $.

The property of the bath is characterized by the spectral distribution function (SDF), defined as
\begin{align}
J_s (\omega) \equiv \sum_{j_s} \frac{\alpha^2_{j_s}}{2 m_{j_s} \omega_{j_s}} \delta (\omega-\omega_{j_s}) .
\label{sec: SDF}
\end{align}
The factor of the counter term is then expressed as
\begin{align}
\label{eq:counter4}
 \Lambda^{(s)}  = \int_0^{\infty} d\omega \frac{J_s(\omega)}{\omega}.
\end{align}

\subsection{HEOM-2DVS}

The noise operator associated with the $s$th intramolecular mode is defined as $\hat X_{s} \equiv \sum_{j_s} \alpha_{j_n} \hat x_{j_s}$. For a harmonic bath, noise correlations beyond the third order do not contribute, and the {\it{dissipation}} can therefore be fully characterized by the linear response function, 
 $iL_1^{(s)} (t) = i\langle {\hat X}_{s}(t) {\hat X}_{s}  -{\hat X}_{s} {\hat X}_{s}  (t) \rangle_B/\hbar$, where $\hat{X}_{s} (t)$ is the Heisenberg representation of $\hat{X}_{s}$ with respect to the bath Hamiltonian $\hat H_B^{(s)}$ (excluding the counter term), and $\langle \cdots \rangle_{B}$ denotes the thermal average over the bath degrees of freedom. Correspondingly, {\it{thermal fluctuations}} are characterized by $L_2^{(s)} (t) =  \langle {\hat X}_{s}(t) {\hat X}_{s} +{\hat X}_{s} {\hat X}_{s}  (t)  \rangle_B/2$. 
The interplay between fluctuation and dissipation facilitates energy exchange, driving the system toward thermal equilibrium. This equilibrium condition is rigorously governed by the quantum fluctuation–dissipation theorem.\cite{T06JPSJ,T20JCP,TK89JPSJ1} The combined kernel function, $L^{(s)} (t) = iL^{(s)} _1 (t) +L^{(s)} _2 (t)$, naturally emerges in the Feynman–Vernon influence functional formalism.\cite{FEYNMAN1963118}

2D IR spectroscopy can directly probe the non‑Markovian nature of the bath through the vibrational dephasing time. The HEOM formalism is capable of accommodating various forms of SDFs;\cite{T20JCP} in this study, we use the simple Drude form:
 \begin{equation}
  J_s(\omega) = \frac{m_s \zeta_s}{2\pi}\frac{\gamma_s^2 \omega}{\omega^2 + \gamma_s^2},
  \label{eq:drude}
  \end{equation}
For the Drude SDF, the dissipation and fluctuation kernels can be expressed as follows:\cite{T06JPSJ,T20JCP} 
\begin{align}
\label{eq:L1}
iL_1^{(s)}(t) &= -\frac{im_s \zeta_s \gamma_s^{2} }{2} {\rm{e}}^{ - {\gamma_s} 
t}
\end{align}
and
\begin{align}
\label{eq:L2}
L_2^{(s)}(t) &=  \frac{m_s \zeta_s \gamma_s^2}{\beta\hbar} \sum_{k=1}^{K_s}\left[\frac{1}{\gamma_s}+\frac{ 2\gamma_s}{\gamma_s^2 - \nu_k^2}\right]\rm{e}^{ - \gamma_s t} 
\nonumber \\ 
& -\frac{m_s \zeta_s \gamma_s^2}{\beta\hbar}\sum_{k=1}^{K_s}\frac{ 2\nu_k}{\gamma_s^2 - \nu_k^2}\rm{e}^{ - \nu_k t},
\end{align}
where $\nu_k = k/\beta \hbar$ are the Matsubara frequencies.

The constant for the counter term is now given by
\begin{equation}
\Lambda^{(s)} = \frac{m_s \zeta_s \gamma_s}{2}.
\label{counterterm}
\end{equation}

For the MBA model [Eqs.\eqref{sec:Total Hamiltonian}-\eqref{eq:h_int}] with the Drude SDF [Eq. \eqref{eq:drude}], \texttt{DHEOM-MLWS} have been formulated to describe two vibrational modes, encompassing both intramolecular and intermolecular dynamics.\cite{TT23JCP1} In parallel, \texttt{CHFPE} have been developed to treat three vibrational modes.\cite{HT25JCP1,HT25JCP2} Hereafter, we refer to this as \texttt{CHFPE-2DVS}. Computational implementations of both approaches are publicly available.\cite{TT23JCP2,HT25JCP2}  

Since this study focuses only on intramolecular vibrational modes, it is feasible to represent the reduced density operator using the eigenenergy states of each vibrational potential rather than phase-space coordinates. It is worth noting that, due to the classical nature induced by thermal baths, the phase-space representation remains advantageous for describing low-frequency intermolecular modes, offering lower computational cost.\cite{
IIT15JCP,IT16JCP,HT25JCP1,HT25JCP2}

The HEOM derived by transforming the phase-space representation into the energy-eigenvalue representation differs from the standard HEOM, as it explicitly includes the counter term in Eq.~\eqref{eq:counter1} as $\hat{H}_{S}'$ in the system Hamiltonian.

To reduce the computational cost, here the function $\coth(x)$ with the Drude SDF is represented
by a meromorphic $[K_s-1/K_s]$ Padé approximant, whose numerator
and denominator have maximum orders $K_s-1$ and $K_s$,
respectively, effectively reducing the contribution of high-order Matsubara terms.
The integer $K_s$ specifies the Padé order for the $s$th mode, and the
resulting Padé-approximated $\nu_k$ and $\eta_k$ are used to construct
the fluctuation and dissipation operators.\cite{hu2010communication} 
We defined $\nu_0^s \equiv \gamma_{s}$, and introduce the Padé approximated frequencies $\nu_k^s$ for $k = \{1, 2, \cdots, K_s\}$.\cite{IT18JCP,IT19JCP} We also define the hyperoperators \(\hat{A}^{\times} \hat{B} \equiv \hat{A} \hat{B} - \hat{B} \hat{A}\) and \(\hat{A}^{\circ} \hat{B} \equiv \hat{A} \hat{B} + \hat{B} \hat{A}\), for arbitrary operators \(\hat{A}\) and \(\hat{B}\).
The HEOM for the Drude SDF is then expressed as\cite{PJT25JCP1}
\begin{eqnarray}
\label{eq:HEOM_DB}
\frac{d}{dt} \hat{\rho}_{\{{\bf n}_s\}} &=& -\left[ \frac{i}{\hbar} \hat{H}_S'^{\times} + \sum_s \sum_{k=0}^{K_s} \left( n_k^s \nu_k^s \right) \right] \hat{\rho}_{\{{\bf n}_s\}} \nonumber \\
&& - i \sum_s \sum_{k=0}^{K_s} n_k^s \hat{\Theta}_k^s \hat{\rho}_{\{{\bf n}_s - {\bf e}_s^k\}} \nonumber \\
&& - i \sum_s \sum_{k=0}^{K_s} \hat{V}_s^{\times} \hat{\rho}_{\{{\bf n}_s + {\bf e}_s^k\}}.
\end{eqnarray}

The hierarchy elements are indexed by the set $\{{\bf n}_s\} \equiv ({\bf n}_1, {\bf n}_{2}, {\bf n}_3)$, where each ${\bf n}_s$ is a multi-index defined as ${\bf n}_s = (n_0^s, n_1^s, \cdots, n_{K_s}^s)$ for the three-mode case. All elements $\hat{\rho}_{\{{\bf n}_s\}}(t)$ with any negative index $n_k^s < 0$ are set to zero.

The notation $\{{\bf n}_s \pm {\bf e}_s^k\}$ indicates an increment or decrement of the $k$th component of ${\bf n}_s$, where ${\bf e}_s^k$ is the unit vector corresponding to the $k$th frequency component in the $s$th bath. The operators are defined as follows:

\begin{eqnarray}
\label{Pade1}
\hat{\Theta}_0^{(s)} &&=- i \frac{m_s \zeta_s\gamma_s^2}{2} \hat{V}_s^{\circ}  \nonumber \\
&&+\frac{m_s \zeta_s\gamma_s}{\beta \hbar} 
\left(1+\sum_{k=1}^{K_s} \frac{2\eta_k^s \gamma_{s}^2 }{\gamma_s^2
-{\nu_k^s}^2}\right)\hat{V}_s^{\times},
\end{eqnarray}
and
\begin{eqnarray}
\label{eq:Pade2}
\hat \Theta_{k> 0}^{(s)}=-\frac{m_s \zeta_s \gamma_s^2 }{\beta \hbar }\frac{2 \eta_k^s \nu _k}{ {\gamma_s^2} - \nu _k^2} \hat{V}_s^\times,
\end{eqnarray}
where the parameters \(\eta_k^s\) denotes the Padé-approximated thermal coupling.\cite{hu2010communication}

\subsection{Linear absorption and 2D correlation IR spectra}
\label{sec:spectruman}

We now examine a model comprising three primary intramolecular modes of the water molecule:  (1) asymmetric stretch, ($1'$) symmetric stretch, and (2) bending.  
These modes are described by dimensionless vibrational coordinates $\bm{q}=(q_1, q_{1'}, q_2)$.\cite{HT25JCP1}

Representing intramolecular modes in terms of energy eigenstates allows for simulations and analyses based on optical Liouville pathways in electronically excited states.\cite{mukamel1999principles}  
Calculating 2D correlation IR spectra within MD or Wigner representations requires additional effort to eliminate contributions from non-rephasing components.\cite{HT08JCP,TT23JCP2,HT25JCP1}  In contrast, energy eigenstate representations enable straightforward evaluation by simply selecting the corresponding optical Liouville paths.\cite{IT06JCP,IT07JPCA} 
However, due to nonlinear interactions between the molecule and the laser field, as well as among vibrational modes, a large number of Liouville pathways need to be considered for an accurate description. To demonstrate this approach, we symbolically represent the three-mode excited states as $\lvert \mathbf{1} \rangle$ and $\lvert \mathbf{2} \rangle$,  as described below.

We first note that the excitation frequencies of the intramolecular modes are much larger than thermal excitations. Therefore, the initial equilibrium state can be safely assumed to be the ground vibrational eigenstate of each mode,
$|{\bf 0}\rangle = |0_1,0_{1'},0_2\rangle$.

We denote the state obtained by applying the dipole operator $\hat{\mu}$ in Eq.~\eqref{eq:mu} to this state once, symbolically, as $\lvert \mathbf{1} \rangle$. 
The state $\lvert \mathbf{1} \rangle$  includes components such as
$\mu^{1'} \lvert 0_{1}, 1_{1'}, 0_{2} \rangle$ and 
$\mu^{12} \lvert 1_{1}, 0_{1'}, 1_{2} \rangle$, , which correspond to single-excitation and double-excitation states, respectively.
However, since the $\mu^{ss'}$ component is smaller than the $\mu^{s}$ component, the double excitation has a negligible effect.

After time $t_{1}$, a second application of the dipole operator to $\lvert \mathbf{1} \rangle$ returns part of the components to the ground state as $\mu^{1'} |0_1,0_{1'},0_2\rangle$ or $\mu^{12}|0_1,0_{1'},0_2\rangle$. 
Simultaneously, it generates higher-excitation contributions, collectively represented as $|{\bf 2}\rangle$, including
$\mu^{1'} |0_1,2_{1'},0_2\rangle$, $\mu^{12}|2_1,0_{1'},0_2\rangle$, and $\mu^{1'2}|1_1,1_{1'},2_2\rangle$. 
During the time evolution $t_{1}$, $t_2$, and $t_3$, excitation or relaxation may occur to various states other than $\lvert \mathbf{1} \rangle$ due to mode--mode interactions characterized by the coupling strength $\Delta_{nn'\,mm'}^{ss'}$ and interactions with the  bath. For short $t_{1}$, however, such contributions are regarded as relatively minor.

\subsubsection{Linear absorption (1D) spectra}

In the density operator representation, the first-order response functions is expressed as\cite{T06JPSJ,T20JCP} 
\begin{eqnarray}
R^{(1)}(t_{1}) =\qty(\frac{i}{\hbar })\mathrm{tr}\qty{\hat{\mu}\mathcal{G}(t_{1})\hat{\mu}^{\times }\hat{\rho }^{\mathrm{eq}}},
\label{eq:R1} 
\end{eqnarray}
where $\hat{\mathcal{G}}(t)\equiv \exp[ -(i/\hbar) \hat H_{tot}^{\times} t]$, which represents the Green's function (Liouvillian propagator) of the system in the absence of a laser interaction, and $\hat{\rho }^{\mathrm{eq}}$ is the equilibrium state.
The Fourier transform of the above $
	I(\omega)=\int^\infty_0 dt R^{(1)}(t) \exp(i \omega t)
$ 
is equivalent to the linear absorption spectrum. 

We evaluate Eqs.~\eqref{eq:R1} in four steps.\cite{T06JPSJ,T20JCP}
\begin{itemize}
  \item Set a factorized temporary initial condition at $t=-t_{eq}$ as  
        $\hat{\rho}_{\{{\bf n}_s ={\bf 0}\}}(-t_{eq}) = |{\bf 0}\rangle \langle {\bf 0}|$.

  \item Propagate the HEOM up to sufficiently long $t_{eq}$ to attain the equilibrium state  
        $\hat{\rho}^{\mathrm{eq}}_{\{{\bf n}_s\}}$.  
        If the vibrational excitation energy is sufficiently higher than the thermal excitation, the state remains equivalent to the temporal initial condition factorized with the bath.

  \item Excite the system at $t_1=0$ by  
        $\hat{\rho}'(0) = \hat{\mu}^{\times}\hat{\rho}^{\mathrm{eq}}$,  
        which yields components in the states $|{\bf 1}\rangle \langle {\bf 0}|$ and $|{\bf 0}\rangle \langle {\bf 1}|$.

  \item Propagate the perturbed hierarchy under the HEOM, Eqs.~\eqref{eq:HEOM_DB}--\eqref{eq:Pade2} up to time $t_1$:  
        $\hat{\rho}'(t_1) = \mathcal{G}(t_1)\hat{\rho}'(0)$.

  \item Response function evaluation:  
        $R^{(1)}(t_{1}) = i \mathrm{tr}\{\hat{\mu}\hat{\rho}'(t_1)\}/\hbar$,  
        and obtain $I(\omega)$ via FFT.
\end{itemize}
 Note that when the response function is described using the density operator, the dipole moment appears time independent because the SL interaction with the thermal bath does not alter its form. However, in the Heisenberg representation, the dipole operator itself carries explicit time dependence, and thus it can be regarded as varying with time--an effect often described as non‑Condon behavior.\cite{Skiner2005nonCondon}

\subsubsection{2D correlation IR spectra}

\begin{figure}[htbp]
  \centering
  \includegraphics[scale=0.35]{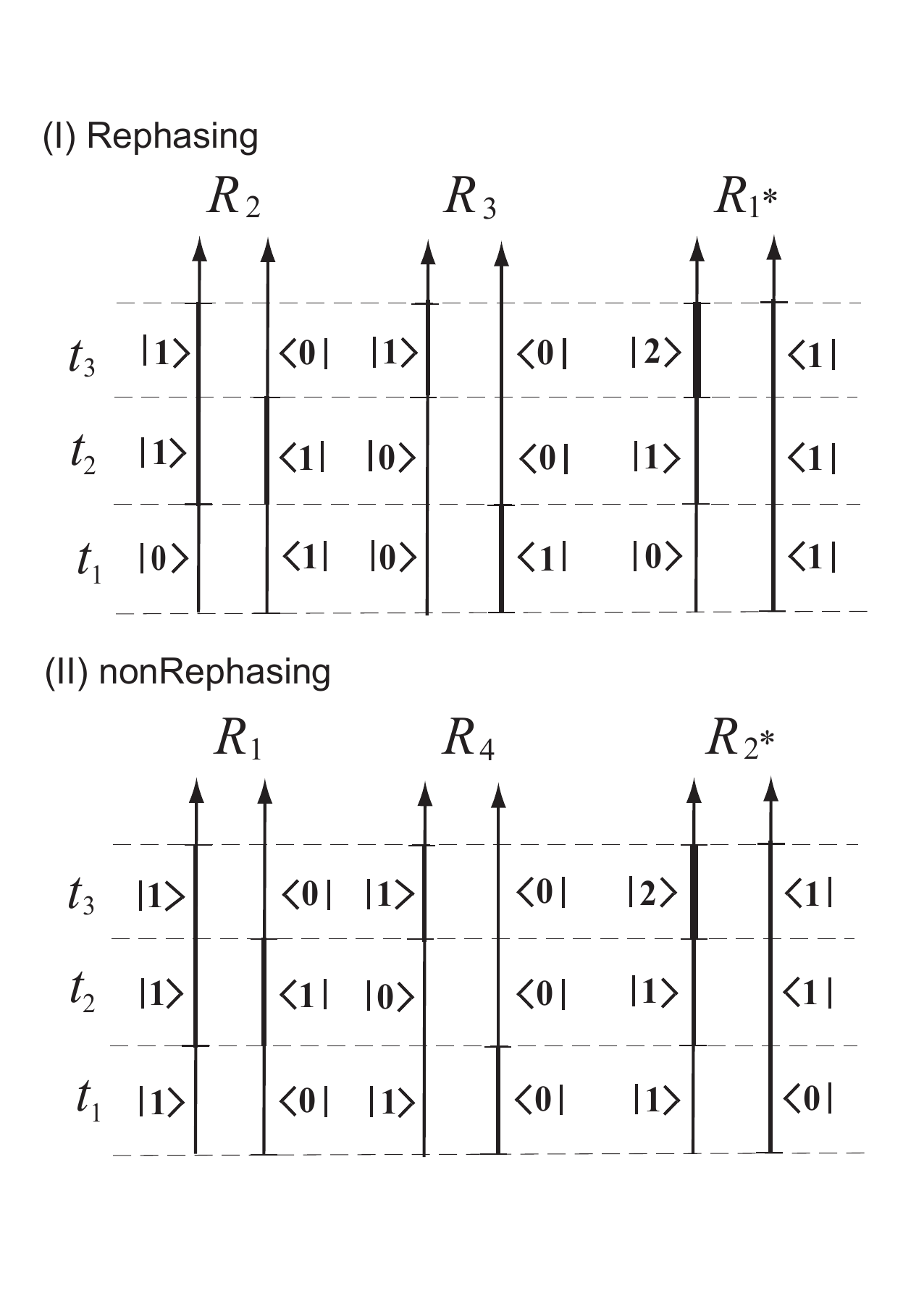}
  \caption{Optical Liouville pathways in 2D vibrational spectroscopy for (I) rephasing and (II) non-rephasing contributions. In each diagram, the left-hand line depicts the time evolution of the ket state $\lvert {\bf n} \rangle$, while the right-hand line depicts that of the bra state  $\langle {\bf n}' \rvert$. The complex-conjugate pathways, obtained by interchanging the left and right states, are not shown.\cite{mukamel1999principles} 
The diagrams were reproduced from Y. Tanimura, J. Chem. Phys. 137, 22A550 (2012), with the permission of AIP Publishing. 
}
  \label{fgr:Liouville}
\end{figure}

For 2DIR experiments, three laser pulses with wavevectors 
$\boldsymbol{k}_1$, $\boldsymbol{k}_2$, and $\boldsymbol{k}_3$ are applied sequentially to the sample at times $0$, $t_1$, and $t_1+t_2$. These pulses generate a four-wave mixing signal field at $t_1+t_2+t_3$ in the phase-matched directions.\cite{mukamel1999principles} 
The signal is described by the third-order nonlinear response function\cite{T06JPSJ,T20JCP}
\begin{eqnarray}
R^{(3)}(t_3,t_2,t_1) = \qty(\frac{i}{\hbar })^{3}\mathrm{tr}\qty{\hat{\mu}\mathcal{G}(t_{3})\hat{\mu}^{\times}\mathcal{G}(t_{2})\hat{\mu}^{\times }\mathcal{G}(t_{1})\hat{\mu}^{\times }\hat{\rho }^{\mathrm{eq}}}.\nonumber \\
  \label{eq:RFTIRI}
\end{eqnarray}
Since $R^{(3)}(t_3,t_2,t_1)$ contains three $\hat{\mu}^{\times}$, the expression consists of eight terms. 

Among them, the rephasing (echo) signal generated along the 
$\boldsymbol{k}_\text{I}=\boldsymbol{k}_3+\boldsymbol{k}_2-\boldsymbol{k}_1$ 
phase-matched direction and the nonrephasing (virtual echo) signal detected along the 
$\boldsymbol{k}_\text{II}=\boldsymbol{k}_3-\boldsymbol{k}_2+\boldsymbol{k}_1$ 
direction are evaluated from\cite{HammLimRobin1998,Demirdoven-Khalil-Tokmakoff:PRL2003,Khalil-Demirdoven-Tokmakoff:JPCA2003,IT06JCP,IT07JPCA,2DCrrJonas2001,2DCrrGe2002,2DCrrTokmakoff2003,T12JCP}
\begin{align}
&R^{(3)}_{\text{I}}(t_3,t_2,t_1)
\notag\\
&\quad \quad = \qty(\frac{i}{\hbar })^{3}
\text{tr}
\left\{\hat{\mu} \mathcal{G}(t_3)\hat{\mu}_\leftarrow^\times \mathcal{G}(t_2)\hat{\mu}_\rightarrow^\times \mathcal{G}(t_1)
\hat{\mu}_\leftarrow^\times \hat{\rho}^\text{eq}\right\},
\label{echo}
\end{align}
and
\begin{align}
&R^{(3)}_{\text{II}} (t_3,t_2,t_1)
\notag\\
&\quad \quad = \qty(\frac{i}{\hbar })^{3}
\text{tr}\left\{\hat{\mu}  \mathcal{G}(t_3)\hat{\mu}_\rightarrow^\times  \mathcal{G}(t_2)
\hat{\mu}_\leftarrow^\times  \mathcal{G}(t_1)
\hat{\mu}_\rightarrow^\times \hat{\rho}^\text{eq}
\right\},
\label{virtual-echo}
\end{align}
respectively, where $\hat{\mu}_\rightarrow \hat A \equiv \hat A \hat{\mu}$ and 
$\hat{\mu}_\leftarrow \hat A \equiv \hat{\mu} \hat A$ for any operator $\hat A$.

By performing the double Fourier transform of Eqs.~\eqref{echo} and \eqref{virtual-echo} 
with respect to $t_1$ and $t_3$, we obtain the 2D rephasing spectrum
\begin{align}
&S_\text{R}(\Omega_3,\Omega_1;t_2)
\notag\\
&\quad \quad=	\text{Im}\iint^\infty_0 dt_3 dt_1 e^{i\Omega_3t_3+i\Omega_1t_1}R^{(3)}_\text{I}(t_3,t_2,t_1),
\label{2D-rephasing}
\end{align}
and 2D nonrephasing spectrum
\begin{align}
&S_\text{NR}(\Omega_3,\Omega_1;t_2)	
\notag\\
&\quad \quad=
\text{Im}
\iint^\infty_0 dt_3 dt_1
e^{i\Omega_3t_3+i\Omega_1t_1}R^{(3)}_\text{II}(t_3,t_2,t_1),
\label{2D-nonrephasing}
\end{align}
respectively.

The individual 2D rephasing and nonrephasing spectra exhibit distorted line shapes 
(phase-twisted lines), because the double Fourier transform mixes absorptive and dispersive features. By adding the rephasing and nonrephasing spectra with equal weights, the dispersive contributions cancel, 
yielding the 2D correlation spectrum with purely absorptive line shapes:\cite{Demirdoven-Khalil-Tokmakoff:PRL2003,Khalil-Demirdoven-Tokmakoff:JPCA2003}
\begin{align}
S_\text{C}(\Omega_3,\Omega_1;t_2)	\equiv S_\text{R}(\Omega_3,-\Omega_1;t_2) + S_\text{NR}(\Omega_3,\Omega_1;t_2).
\label{correlation-spectrum}
\end{align}
 
The procedure for calculating the 2D correlation spectrum 
using Eqs.~\eqref{eq:HEOM_DB}-\eqref{eq:Pade2} is summarized below, 
taking the rephasing contribution $R_{2}$ in Fig.~\ref{fgr:Liouville}(I) as an illustrative example.
\begin{itemize}
  \item Initial condition: The method for setting the initial conditions $\hat{\rho}^{\mathrm{eq}}_{\{{\bf n}_s\}}=\lvert {\bf 0} \rangle \langle {\bf 0} \rvert$ is the same as for linear absorption.    
  \item First interaction ($t=0$):  
    $\hat{\rho}'(0) = \hat{\rho}^{\mathrm{eq}}\hat{\mu}$, yielding $|{\bf 0}\rangle \langle {\bf 1}|$.  
    Propagation under HEOM up to $t_1$:  
    $\hat{\rho}'(t_1) = \mathcal{G}(t_1)\hat{\rho}'(0)$.

  \item Second interaction ($t_1$):  
    $\hat{\rho}''(t_1) = \hat{\mu}\hat{\rho}'(t_1)$.  
    Propagation up to $t_1+t_2$:  
    $\hat{\rho}''(t_1+t_2) = \mathcal{G}(t_2)\hat{\rho}''(t_1)$.

  \item Third interaction ($t_1+t_2$):  
    $\hat{\rho}'''(t_1+t_2) = \hat{\mu}\hat{\rho}'(t_1+t_2)$.  
    Propagation up to $t_1+t_2+t_3$:  
    $\hat{\rho}'''(t_1+t_2+t_3) = \mathcal{G}(t_3)\hat{\rho}'''(t_1+t_2)$.

  \item Response function evaluation:  
    $R_2 (t_1,t_2,t_3) = \langle {\bf 0}|\hat{\mu}\,\hat{\rho}'''(t_1+t_2+t_3)|{\bf 0}\rangle$ and $\langle {\bf 2}|\hat{\mu}\,\hat{\rho}'''(t_1+t_2+t_3)|{\bf 2}\rangle$. 
\end{itemize}

The contributions from the other diagrams can also be calculated in the same manner. Contributions containing only population states are referred to as rephasing part at $t_{2}$ intervals, whereas contributions involving other coherences are termed non-rephasing part. 

Note that the vibrational modes expressed in the energy eigenstate representation allow the rephasing contribution to be isolated through the Liouville-space pathways, as described in this work. In contrast, for systems formulated in the classical phase space or in the quantum Wigner representation, the rephasing component can be extracted by performing the Fourier transforms with respect to $t_1$, $t_2$, and $t_3$.\cite{TI09ACR,TT23JCP1,TT23JCP2,HT25JCP1,HT25JCP2,Khalil-Demirdoven-Tokmakoff:JPCA2003,IT06JCP,IT07JPCA,2DCrrJonas2001,2DCrrGe2002,2DCrrTokmakoff2003,T12JCP} 

\section{Computational Framework for HEOM-2DVS}
\label{sec:HEOMcode} 

Numerical integration of the HEOM has enabled detailed analyses of diverse open quantum dynamical problems.
Consequently, many practical implementations have been developed.\cite{Shi2009HEOM,SHI10.1063/1.5026753,
IKEDASCHOLES2020,IKEDANAKAYAMA2022,Huang2023,Ke2023TreeTensorHEOM,Borrelli2021,Takahashi2024,Ignacio2024,Ignacio2025,LinjunWan2025}  
The \texttt{HEOM-2DVS} employed in this study differs from conventional formulations by explicitly incorporating counter terms from the MAB model and handling three independent thermal baths with coupled LL+SL interactions. Because intramolecular vibrational excitation energies exceed thermal energies, a larger low-temperature correction term is required. Furthermore, evaluating the 2D correlation spectrum requires scanning $t_{1}$ and $t_{3}$ for each value of $t_{2}$ and performing Fourier transforms, resulting in substantial computational cost. Thus, reducing computational cost is essential for applying HEOM to the MAB.

The model employed in this simulation is identical to the one used in our previous classical calculations.\cite{HT25JCP1} Calculations based on the Wigner distribution function are well suited for studying intermolecular vibrations whose excitation energies are close to the thermal energy,\cite{HT25JCP2} whereas solving the HEOM in the energy‑eigenstate representation is numerically more efficient for quantum treatments of intramolecular vibrations with much higher excitation energies. Thus, the HEOM for MAB and the corresponding QHFEP provide complementary capabilities. Accordingly, the input for the HEOM used in this code was standardized to match the format of the HFPE. Specifically, the inputs consist of the parameter values appearing in Eqs.\eqref{sec:Total Hamiltonian}–\eqref{eq:h_int}. 

The time evolution of the HEOM was computed using the Runge--Kutta method with a time step of $dt = 0.15\,\mathrm{fs}$. When integrating the HEOM, we varied the number of hierarchy levels (i.e., the number of Padé terms) to ensure convergence. For the four-level system considered here, we used $K_1 = 4$, $K_{1'} = 4$, and $K_2 = 4$.
The implementation efficiently manages large queues through external libraries including those with GPU support.
Additional details can be found in the accompanying  \texttt{README.pdf} file.

Numerical calculations were performed on two PCs:
(i) Intel(R) Core(TM) i9-9900K 8-core CPU with 32 GB RAM and an NVIDIA GeForce RTX 3080 Ti (12 GB), and
(ii) Intel(R) Xeon(R) Gold 6212U CPU with 192 GB RAM and an NVIDIA A100 GPU (40 GB).
The operating system was AlmaLinux 8, the compiler was GCC 13, and Python 3.12 was used.
The C++/CUDA libraries included Eigen, HDF5 1.14.6, the CUDA runtime, cuBLAS, and cuSPARSE.
The Python libraries used were NumPy, Matplotlib, h5py, and pybind11.

On the A100 GPU, for a three vibrational-mode system with four energy levels, using a time step of $dt = 0.25\,\mathrm{fs}$ and 33,826 hierarchy elements, the total computation time required to obtain all linear-response signals for $t_2 = 0\,\mathrm{fs}$, $50\,\mathrm{fs}$, and $100\,\mathrm{fs}$ in a single run was approximately $7.5 \times 10^4\,\mathrm{s}$.
The corresponding GPU memory (VRAM) consumption was approximately 1.5 GB.

\section{Numerical Demonstration}
\label{sec:Demo}

\begin{table*}[!tb]
 \caption{\label{tab:para2mode}Parameter values for (i) the two-mode MAB model consisting of the (1) stretching mode and (2) bending mode. 
The values were taken from Refs.~\onlinecite{TT23JCP1,TT23JCP2}. The fundamental frequency was set to $\omega_{0} = 4000~\mathrm{cm}^{-1}$. 
The anharmonic mode-mode coupling and dipole elements are $\tilde{g}_{1^{2}1'} = 0$, $\tilde{g}_{11'^{2}} = 0.2$, and $\tilde{\mu}_{11'} = 2.0 \times 10^{-3}$. 
The normalization parameters are defined as 
$\tilde{\zeta}_s \equiv (\omega_0/\omega_s)^2 \zeta_s$, 
$\tilde{V}_{LL}^{(s)} \equiv (\nu_s/\nu_0) V_{LL}^{(s)}$, 
$\tilde{V}_{SL}^{(s)} \equiv V_{SL}^{(s)}$, 
$\tilde{g}_{s^{3}} \equiv (\nu_s/\nu_0)^3 g_{s^{3}}$, 
$\tilde{\mu}_{s} \equiv (\mu_0/\omega_s)\mu_{s}$, 
and $\tilde{\mu}_{ss} \equiv (\nu_0/\omega_s)^2 \mu_{ss}$.%
The table was reproduced from H.
Takahashi and Y. Tanimura, J. Chem. Phys. 158, 124108
(2023), with the permission of AIP Publishing.
}
\scalebox{0.9}{
\begin{tabular}{ccccccccc}
  \hline \hline
s&    $\nu_s$ (cm$^{-1}$) & $\gamma_s/\omega_0$ & $\tilde{\zeta}_s$ & $\tilde{V}_{LL}^{(s)}$ & $\tilde{V}_{SL}^{(s)}$ & $\tilde{g}_{s^3}$ & $\tilde{\mu}_{s}$  & $\tilde{\mu}_{ss}$\\
  \hline
1&   $3520$ & $5.0{\times}10^{-3}$ & $9$ & $ 0 $                     & $1.0$ & $-5.0{\times}10^{-1}$ & $ 3.3 $  & $1.2{\times}10^{-2}$\\
2&  $1710$ & $2{\times}10^{-2}$ & $0.8$ & $ 0 $                   & $1.0$ & $-7{\times}10^{-1}$ & $ 1.8 $ & $0$\\
  \hline
\end{tabular}
}
\end{table*}

To demonstrate the capabilities of the \texttt{HEOM‑2DVS}  framework, we calculated the linear absorption (1DIR) spectra and the 2D correlated IR spectra for the intramolecular vibrational modes of liquid water using both (i) two-mode and (ii) three-mode models.

Quantum calculations were carried out using both three‑level and four‑level representations for each vibrational mode, and the results were compared. The 1D IR spectra were nearly identical for the two representations, while subtle differences emerged in the 2D IR spectral profiles, with the four-level representation providing an enhanced description.
The results obtained using the four-level eigenstate representation
from $|0_s\rangle$ to $|3_s\rangle$ for mode $s$ are shown below for the two-mode
and three-mode cases.

For (i) the two-mode calculations, the fundamental frequencies were set to (1) stretch mode ($\omega_1 = 3520~\mathrm{cm}^{-1}$) 
and (2) bend mode ($\omega_2 = 1710~\mathrm{cm}^{-1}$) to enable direct comparison with the \texttt{DHEOM-MLWS} results obtained in the Wigner-space 
representation.\cite{TT23JCP2}  The parameters adopted here were chosen to match those used in that calculation and are summarized in Table~\ref{tab:para2mode}.

For (ii) the three-mode calculations, the fundamental frequencies were set to (1) asymmetric stretch ($\omega_1 = 3570~\mathrm{cm}^{-1}$), (1$'$) symmetric stretch ($\omega_{1'} = 3470~\mathrm{cm}^{-1}$), and (2) bending ($\omega_2 = 1710~\mathrm{cm}^{-1}$) and employ the parmeter vaules for the strong intermolecular-coupling case in classical 
simulations based on the \texttt{CHFPE-2DVS} framework.\cite{HT25JCP1}  The parameter are summarized in Tables~\ref{tab:para1} and \ref{tab:FitAll2}.

We note that the symmetric and asymmetric-stretching modes are not distinguishable in the 2DIR spectrum. As a result, in the three-mode case, their individual parameters and mutual coupling cannot be uniquely identified based solely on the 2D spectral data.\cite{IT16JCP}  Thus, the intermode‑couplings presented in Table \ref{tab:FitAll2} is treated not as a fixed parameter but as an adjustable variable used to fit the MD simulation results and experimental results.

\begin{table*}[!tb]
\caption{\label{tab:para1}
Parameter values of (ii) the three-mode MAB model for the (1) asymmetric stretching, (1$'$) symmetric stretching, and (2) bending modes. To enable comparison with the classical results, the parameters were set to the same values as those used in the three-mode \texttt{CHFPE-2DVS} calculation in Ref.~\onlinecite{HT25JCP1}. 
The fundamental frequency was fixed at $\omega_{0} = 4000~\mathrm{cm}^{-1}$. 
The intermode coupling strengths and optical properties are given in Table~\ref{tab:FitAll2}.
The table was reproduced from R.
Hoshino and Y. Tanimura, J. Chem. Phys. 162, 044105
(2025), with the permission of AIP Publishing. 
}
\scalebox{0.9}{
\begin{tabular}{cccccccccc}
  \hline \hline
s&    $\nu_s$ (cm$^{-1}$) & $\gamma_s/\omega_0$ & $\tilde{\zeta}_s$ & $\tilde{V}_{LL}^{(s)}$ & $\tilde{V}_{SL}^{(s)}$ & $\tilde{g}_{s^3}$ & $\tilde{\mu}_{s}$  & $\tilde{\mu}_{ss}$\\
  \hline
1&   $3570$ & $5.0{\times}10^{-3}$ & $9$ & $ 0 $                     & $1.0$ & $-5.0{\times}10^{-1}$ & $ 3.3 $ & $1.2{\times}10^{-2}$\\
1'&   $3470$ & $5.0{\times}10^{-3}$ & $9$ & $ 0 $                     & $1.0$ & $-5.0{\times}10^{-1}$ & $ 3.3 $& $1.2{\times}10^{-2}$\\
2&  $1710$ & $2{\times}10^{-2}$ & $0.8$ & $ 0 $                   & $1.0$ & $-7{\times}10^{-1}$ & $ 1.8 $& $0$\\
  \hline
  \hline \hline\\
\end{tabular}
}
\end{table*}

\begin{table*}[!tb]
\caption{\label{tab:FitAll2}
Parameter values of the anharmonic intermode couplings and optical properties for (ii) the three-mode MAB model (see also Table~\ref{tab:para1}). The intermolecular coupling strength employed here corresponds to the strong-coupling case in Ref.~\onlinecite{HT25JCP1}.
The table was reproduced from R.
Hoshino and Y. Tanimura, J. Chem. Phys. 162, 044105
(2025), with the permission of AIP Publishing. 
}
\begin{tabular}{ccccccccccc}
  \hline \hline
  $\mathrm{s-s'}$ & $\tilde{g}_{ss'}$  & $\tilde{g}_{s^2s'}$ & $\tilde{g}_{s{s'}^2}$ & $\tilde{\mu}_{ss'}$ \\
  \hline
  $\mathrm{1-1'}$  & $-5{\times}10^{-3}$ & $0.32$ & $-4.2{\times}10^{-3}$ & $ 0$ \\
  $\mathrm{1-2}$  & $1{\times}10^{-7}$ & $-2,6{\times}10^{-2}$ & $4{\times}10^{-4}$ & $2.0 \times 10^{-3}$ \\
  $\mathrm{1'-2}$  &  $-8{\times}10^{-4}$ &$1.2{\times}10^{-1}$ & $-1.2{\times}10^{-2}$ & $2.0 \times 10^{-3}$ \\
  \hline \hline\\
\end{tabular}
\end{table*}

\subsection{Linear Absorption (1D) spectra}
\begin{figure}[htbp]
  \centering
  \includegraphics[keepaspectratio, scale=0.45 ]{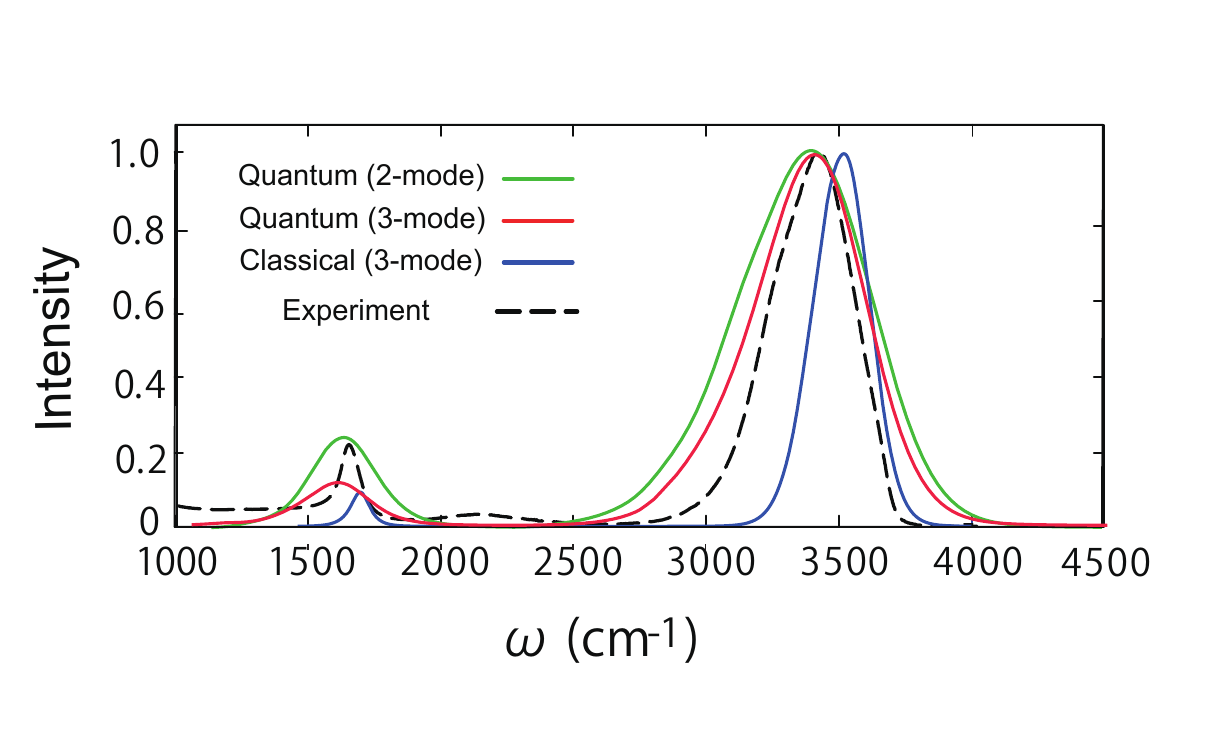}
  \caption{Linear absorption (1DIR) spectrum of water calculated for the two‑mode and three‑mode MAB models using \texttt{HEOM‑2DVS} (quantum) and \texttt{CHFPE-2DVS} (classical). The experimental IR data  are shown as dashed black curves for comparison. Each spectrum is normalized to its maximum peak intensity. The blue solid curves represent the three‑mode classical result, while the green and red solid curves represent the two‑mode and three‑mode quantum results, respectively.
The H$_2$O experimental spectrum is reproduced
from J. Chem. Phys. 131, 184505 (2009), with the permission of AIP Publishing.\cite{D2OIR} 
}
  \label{fgr:linear}
\end{figure}

The 1DIR spectra for the quantum two-mode case (Table~\ref{tab:para2mode}) 
and for the classical and quantum three-mode cases 
(Tables~\ref{tab:para1} and \ref{tab:FitAll2}) are shown in 
Fig.~\ref{fgr:linear}. The quantum two-mode and three-mode spectra were 
computed using \texttt{HEOM-2DVS}, whereas the classical three-mode spectrum was obtained 
using \texttt{CHFPE-2DVS}.\cite{HT25JCP1}

In IR spectra, the stretching and bending peaks of high-frequency intramolecular modes appear blue-shifted in classical descriptions because quantum anharmonic effects are absent in such treatments.\cite{ST11JPCA}

Although the present MAB model was constructed from classical 2D IR–Raman simulations, the underlying 
force field (\texttt{POLI2VS})\cite{HT11JPCB} was originally developed for quantum MD. Consequently, even though the MAB model is derived from classical MD 
trajectories, it yields vibrational spectra that remain accurate when combined with quantum HEOM calculations, producing results comparable to 
quantum MD simulations using \texttt{POLI2VS}.\cite{JianLiu2018H2OMP} This suggests 
that a quantum MAB model can, in principle, be extracted from 2D spectra 
generated by first-principles classical MD simulations in which nuclear 
motion is treated classically.\cite{TT23JCP1,TT23JCP2} 

Besides the blue-shifted peak positions, the classical spectra show narrower linewidths compared to the quantum-mechanically calculated spectra. This is because, in the classical case, nuclear wave packets are confined near the bottom of the potential, while in the quantum case, they spread out due to zero-point vibrations. The resulting broadening of the wave packets leads to wider IR linewidths in the quantum simulations.

As shown here, the energy‑eigenstate representation provides an efficient framework for capturing quantum effects involving three intramolecular modes at relatively low computational cost. It also serves as a useful complement to \texttt{DHEOM‑MLWS}, which is currently limited to two‑mode systems.

\subsection{2D Correlation IR Spectra}

We now present the 2D correlation IR spectra for the two‑mode and three‑mode cases. The 2D results for the classical description are shown in Ref.~\onlinecite{HT25JCP1}.

\subsubsection{2 Modes (one stretch and one bend) case}

\begin{figure}[htbp]
  \centering
  \includegraphics[keepaspectratio, scale=0.45]{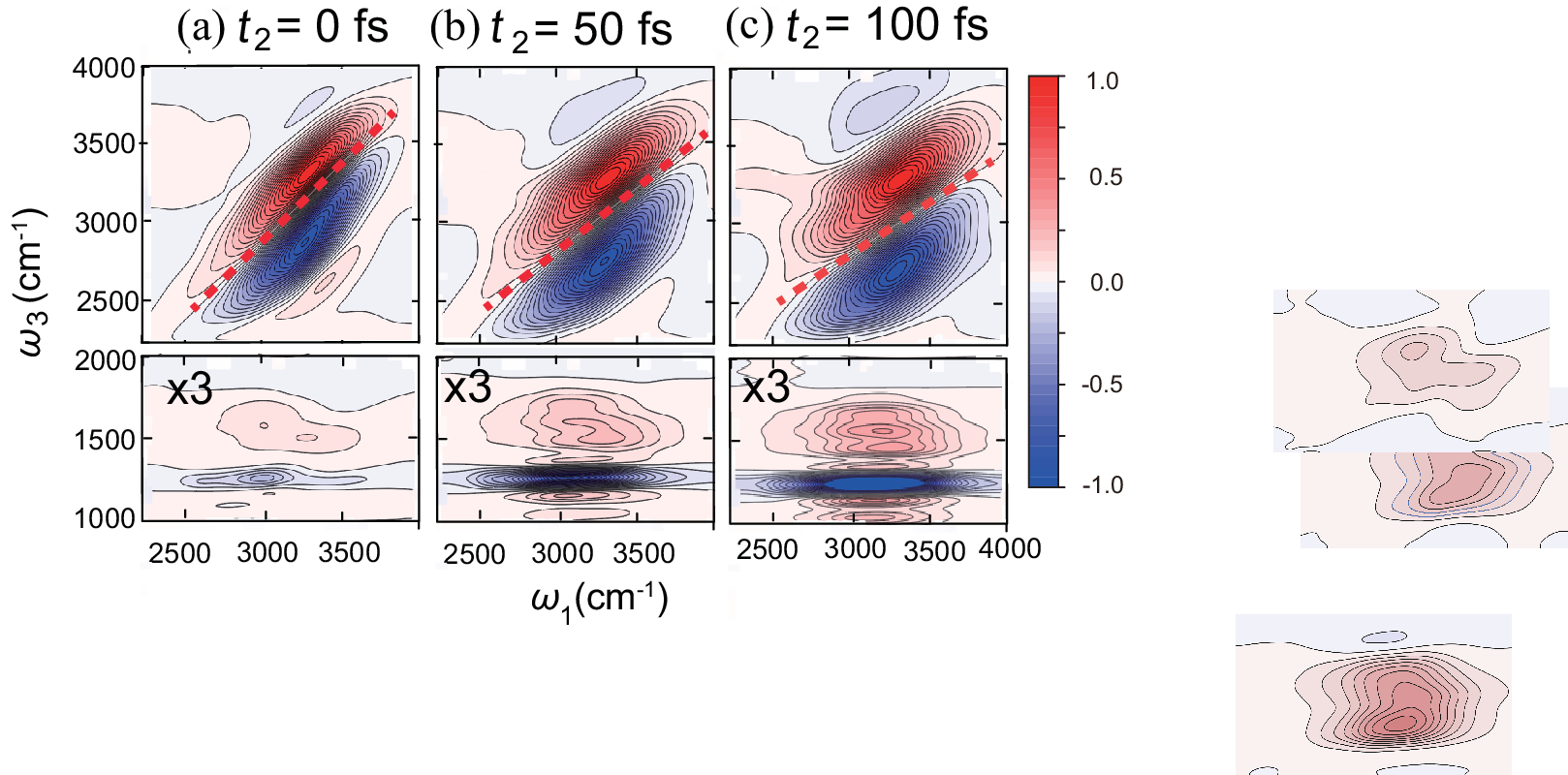}
  \caption{2D correlation IR spectra for the stretching (upper panel) and 
stretching$\rightarrow$bending motions (lower panel) calculated using the two-mode model, 
which includes (1) the OH stretching mode ($\omega_1 = 3520$ cm$^{-1}$), 
and (2) the HOH bending mode ($\omega_2 = 1710$ cm$^{-1}$). 
Spectral intensities were normalized to the maximum amplitude of streching mode.
Because the peak intensity of the lower panels is weaker than in the upper panel, the contour interval was tripled for clarity.
}
  \label{fgr:2DIR2mode_s}
\end{figure}

\begin{figure}[htbp]
  \centering
  \includegraphics[keepaspectratio, scale=0.45]{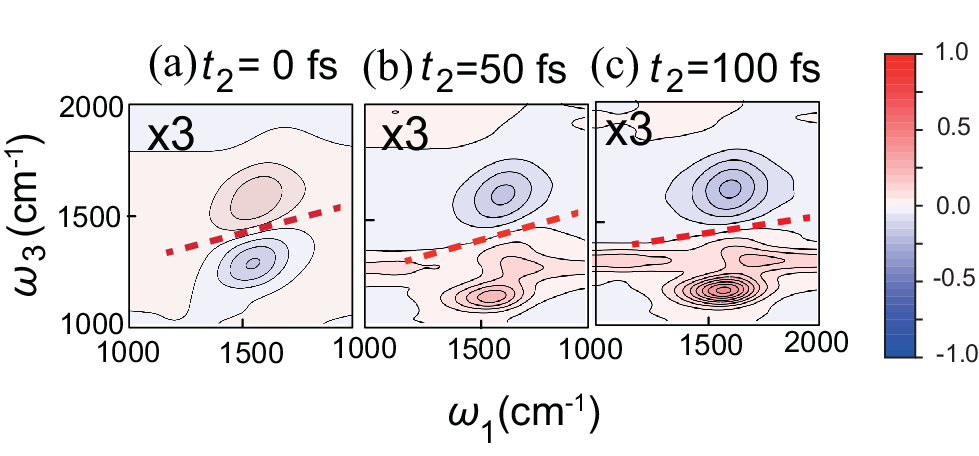}
  \caption{2D correlation IR spectra for the bending motion for the two-mode case. As the peak intensity was weaker than that in the upper panel of Fig. \ref{fgr:2DIR2mode_s}, the contour interval was tripled for emphasis.}
  \label{fgr:2DIR2mode_b}
\end{figure}

We first present the results for the two‑mode case and discuss the differences in the description that appear in the \texttt{DHEOM‑MLWS} results.\cite{TT23JCP2}

Figure~\ref{fgr:2DIR2mode_s} illustrate the 2D correlation IR spectra calculated for the stretching--bending (1--2) modes. 
The upper panel shows the stretching-mode peaks near $(\omega_1,\omega_3) = (3400~\mathrm{cm}^{-1}, 3400~\mathrm{cm}^{-1})$, where the red positive and blue negative features arise from the 
$|0_s\rangle \!\rightarrow\! |1_s\rangle \!\rightarrow\! |0_s\rangle$ and 
$|0_s\rangle \!\rightarrow\! |1_s\rangle \!\rightarrow\! |2_s\rangle$ pathways for $s=1$, respectively, with $|n_s\rangle$ denoting the $n$th vibrational eigenstate of mode $s$.

The orientation of the red dashed nodal lines reflects the degree of noise correlation (non-Markovian effects) between the vibrational coherences during $t_1$ and $t_3$. 
A direction parallel to the $\omega_1$ axis corresponds to the uncorrelated limit, whereas alignment along the $\omega_1 = \omega_3$ diagonal corresponds to the fully correlated limit.\cite{2DCrrJonas2001,2DCrrGe2002,2DCrrTokmakoff2003} 
The peak width parallel to the $\omega_1 = \omega_3$ line reflects inhomogeneous broadening, while the width perpendicular to this line reflects homogeneous broadening.\cite{Wiersma2006,TI09ACR} 

Compared with the 2D spectra calculated using only three energy eigenstates (not shown), an elongated red feature develops near $\omega_{3} \approx 3200\ \mathrm{cm^{-1}}$, while a slightly elongated blue feature emerges around $2800\ \mathrm{cm^{-1}}$ as $t_{2}$ increases.
These features are assigned to contributions from the $\lvert 3_1 \rangle$ state. 
In contrast, no discernible contribution from $\lvert 3_1 \rangle$ is observed at $t_{2}=0$.

The lower panel in Fig.~\ref{fgr:2DIR2mode_s} displays the cross peaks associated with the stretching$\rightarrow$bending transition (e.g., 
$|0_1\rangle|0_2\rangle \!\rightarrow\! |1_1\rangle|0_2\rangle 
\sim 
|0_1\rangle|1_2\rangle \!\rightarrow\! |0_1\rangle|0_2\rangle$).\cite{TI09ACR} 
The coupling peak observed at $t_2 = 0$ arises from coherent energy exchange between the two modes. 
While this coherent peak decays rapidly, the peak appearing around $t_2 = 50~\mathrm{fs}$ reflects population transfer, and its intensity increases with increasing $t_2$.\cite{TT23JCP2} 
The involvement of the $\lvert 3_1 \rangle$ transition leads to a more complex peak profile once $t_2$ exceeds 50 fs.

The  \texttt{HEOM-2DVS} results are qualitatively similar to those of \texttt{DHEOM-MLWS}, with the exception that the diagonal (inhomogeneous) broadening of the two peaks extends over a wider range (2600–3800 $\mathrm{cm}^{-1}$) compared to 3200–3700 $\mathrm{cm}^{-1}$, while the off-diagonal (homogeneous) broadening is narrower
.\cite{TI09ACR} 
As discussed in the 1D spectrum, this broadening is attributed to the strong quantum character of the enhanced zero-point vibrations. 
Because this broadening reflects coherence, the inhomogeneity manifested as off-diagonal broadening remains small.

The 2D correlation IR spectra for the bending mode are presented in Fig.~\ref{fgr:2DIR2mode_b}.
The \texttt{DHEOM-MLWS} results\cite{TT23JCP2} show that the nodal line is initially horizontal, 
whereas the \texttt{HEOM-2DVS} results exhibit strong coherence, leading to a clear relaxation of the nodal line.

For $t_{2} \ge 50\ \mathrm{fs}$, the red peak exhibits a clear splitting. 
A comparison with the spectra obtained from the three-eigenstate calculation (not shown) indicates that this splitting originates from transitions involving the 
$\lvert 3_2 \rangle$ state.

The more pronounced quantum effects observed in \texttt{HEOM‑2DVS}, compared to \texttt{DHEOM‑MLWS} , are likely due to the limitation of each vibrational mode to four energy eigenstates.
In particular, the low‑frequency bending mode is expected to require a larger number of eigenstates for an accurate description.

\subsubsection{3 Modes (two stretches and one bend)  case}

\begin{figure}[htbp]
  \centering
  \includegraphics[keepaspectratio, scale=0.45]{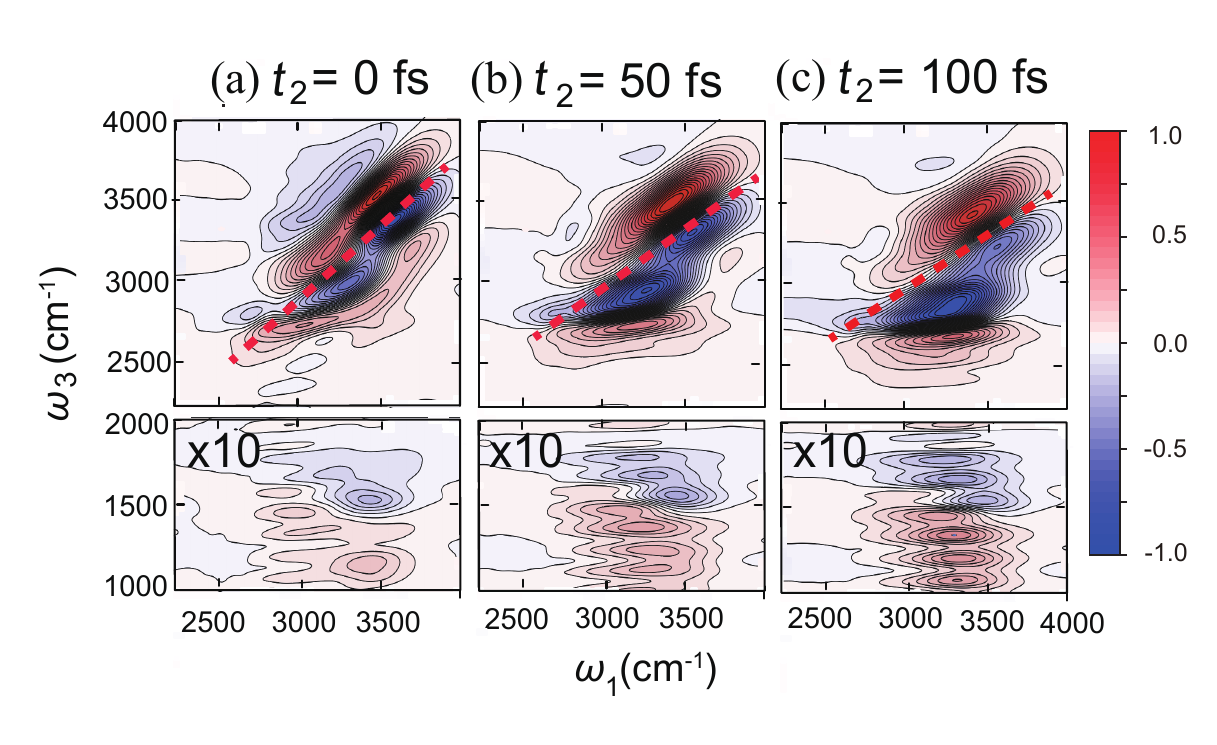}
  \caption{2D correlation IR spectra for the stretching  (upper panel) and 
stretching$\rightarrow$bending motions  (lower panel) calculated using the three-mode model, 
which includes (1) the OH stretching mode ($\omega_1 = 3570$ cm$^{-1}$), 
($1'$) the OH asymmetric-stretching mode ($\omega_{1'} = 3470$ cm$^{-1}$), 
and (2) the HOH bending mode ($\omega_2 = 1710$ cm$^{-1}$). 
The mode-mode coupling among the three modes was set to the strong-coupling 
values listed in Table~\ref{tab:FitAll2}. 
Spectral intensities were normalized to the maximum amplitude of strech peak.
The contour interval was increased by a factor of ten for emphasis.
}
  \label{fgr:2D3Modelweak}
\end{figure}

\begin{figure}[htbp]
  \centering
  \includegraphics[keepaspectratio, scale=0.45]{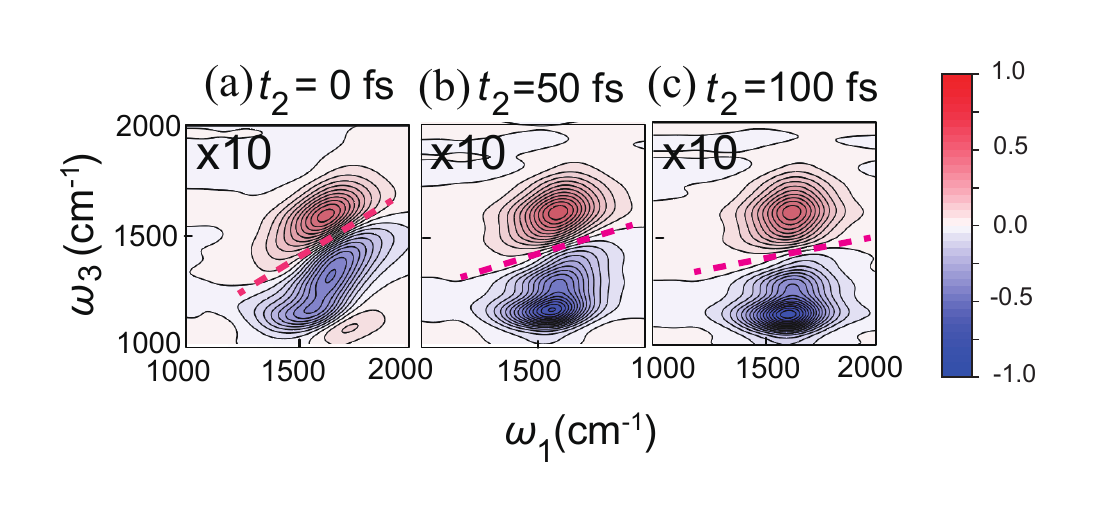}
  \caption{Results for the same calculations as in Fig.~\ref{fgr:2D3Modelweak}, but 
for the bending modes of the three-mode model. The 
contour interval was increased by a factor of ten for clarity.
}
  \label{fgr:2D3ModeBendWeak}
\end{figure}

Figures~\ref{fgr:2D3Modelweak} shows the 2D correlation spectra calculated for 
the OH stretching mode with $\omega_1$=3570 cm$^{-1}$ and $\omega_{1'}$=3470 cm$^{-1}$(upper panel), and the stretching$\rightarrow$bending cross peaks (lower panel), while Fig. \ref{fgr:2D3ModeBendWeak} presents the corresponding HOH bending mode at 
$\omega_2$=1710 cm$^{-1}$.\cite{Tokmakoff2016H2O,VothTokmakoff_St-BendJCP2017,Tokmakoff2022} 

Compared with the classically obtained 2D spectrum calculated using the same parameters,\cite{HT25JCP1} the present quantum results exhibit clearly separated peaks for the symmetric and asymmetric stretching modes. 
This distinction originates from peak splitting caused by strong coupling between the inter‑stretch modes, together with the enhanced ability of quantum computing approaches to preserve vibrational coherence.
The asymmetric stretch is strongly coupled to the bending motion, and as its coherence decays rapidly, the corresponding asymmetric–bending cross peak becomes increasingly prominent.

The stretching$\rightarrow$bending cross peaks  in the lower pannel of Fig.~\ref{fgr:2D3Modelweak} and bending peaks in Fig. \ref{fgr:2D3ModeBendWeak} are much weaker than in the two-mode case as we observed in 1DIR spectra. Their profiles are elongated only along $\omega_1$, reflecting identical anharmonicities of the two stretching modes. The peak intensities increase with $t_2$ because of population transfer. 
Due to the contributions from the transitions originating from $\lvert 3_{1}\rangle$ and $\lvert 3_{1'}\rangle$, numerous parallel peaks emerge as $t_{2}$ increases.

Regarding the two stretching modes, the vibrational states are defined as
$ |{\mathbf{1}}^{\pm}\rangle = |{\bf 0}\rangle = |1_1,0_{1'},0_2\rangle 
  \pm  |0_1,1_{1'},0_2\rangle 
)/\sqrt{2}$, and $|{\mathbf{2}}^{\pm}\rangle
=  ( |2_1,0_{1'},0_2\rangle 
  \pm  |0_1,2_{1'},0_2\rangle )/\sqrt{2}$. 
Then the large red and blue peaks can be assigned to the $|\bm 0\rangle \rightarrow |{\bm 1}^{\pm}\rangle \rightarrow |\bm 0\rangle$ and $|\bm 0\rangle \rightarrow |{\bm 1}^{-}\rangle \rightarrow |{\bm 2}^- \rangle$ pathways, respectively, whereas the third blue peak originates from the $|\bm 0\rangle \rightarrow |{\bm 1}^{\pm}\rangle \rightarrow |{\bm 2}^+\rangle$ transition. These transitions proceed through coherent dynamics, with the associated peaks appearing immediately at $t_{2}=0$ and decaying as $t_{2}$ increases.

The main difference between the 2 mode and 3 mode results—aside from the emergence of symmetric and asymmetric patterns in the 2D IR spectra—is the appearance, in the 3 mode calculations, of a third blue peak on the high‑frequency side, located above the red peak.
Although this third peak is not observed experimentally, it can be attributed to the strong intermode coupling between the symmetric and asymmetric‑stretch modes, 
which was treated as an adjustable parameter in the present model. Consequently, this peak does not appear in the 2 mode case.

\section{Conclusion}
\label{sec:conclude}

In this work, we addressed the long-standing challenge of simulating 2D vibrational spectra in solution, particularly for intramolecular modes whose molecular motions are quantum-mechanically entangled with their environment. To date, no simulation has successfully reproduced the 2D spectra of the intramolecular vibrations of water while combining MD with a quantitatively accurate description of quantum dissipation.

Within these limitations, the MAB model provides a highly descriptive framework capable of reproducing experimental features while incorporating complex intermolecular interactions and the nonlinear system–bath couplings responsible for vibrational dephasing. Analyzing 2D signals using this model-based approach provides clarity on the physical origins of spectral line-shape features and offers conceptual insights that are challenging to derive from fully detailed MD simulations.
The current code facilitates rapid computation of 2D signals by modeling each intramolecular vibration as a four-level system, albeit with less descriptive power compared to \texttt{DHEOM-MLWS}.\cite{TT23JCP1,TT23JCP2}

The model parameters used in this study were selected to replicate the 2D IR–Raman signals derived from MD simulations. In conjunction with this study, we conducted a separate investigation where the parameters of the same MAB model were optimized using ML techniques derived from MD trajectories.\cite{PJT25JCP1,PUT26JCP1} These findings suggest that the representativeness of the underlying MD trajectories, which is heavily influenced by the choice of MD potential, plays a significant role in determining the 2D spectra obtained in this study.

In this paper, we therefore limit ourselves to demonstration calculations rather than pursuing a detailed analysis. The primary objective is to provide a numerical program, and the numerical procedures used here follow those of our previous studies. Although the resulting 2D signal profiles are reasonable, a detailed comparison with experimental data and with simulations employing alternative potential models remains an important direction for future work. In a subsequent study, we will apply HEOM parameters constructed from MD trajectories using ML techniques to analyze the 2D spectra of H$_2$O and D$_2$O and we will discuss the distinct underlying physical processes revealed through their IR spectra.\cite{PHT26JCP2}

The source codes provided here complement the capabilities of \texttt{DHEOM-MLWS},\cite{TT23JCP1,TT23JCP2} which treats two modes quantum mechanically, and  \texttt{CHFPE-2DVS},\cite{HT25JCP1,HT25JCP2}  which treats three modes classically, while enabling nonlinear spectral calculations within a non‑perturbative and non‑Markovian thermal‑bath framework.  A natural future extension of \texttt{HEOM‑2DVS} would be to incorporate intramolecular modes that can be treated classically. This would enable us to investigate the flow of energy and phase from intramolecular to intermolecular modes.

Building on this foundation, the present time‑evolution engine supports numerically ``exact'' simulations of three‑site quantum systems possessing four or more energy levels, each interacting with an independent thermal bath. The framework thus offers a flexible basis for future developments, including applications to electron, exciton, and proton transport.\cite{ZBT20JCP,ZBT21JCP,NT21JCP,CT21JCP,CBT22JCP}

\section*{Supplementary Material}
Numerical integration codes on the basis of HEOM formalism for 1D IR and 2D correlation IR (\texttt{HEOM-2DVS}) are provided as supplemental materials. The manual can be found in the ReadMe.pdf file.

\section*{Acknowledgments}
Y. T. was supported by JST (Grant No. CREST 1002405000170).  

\section*{Author declarations}
\subsection*{Conflict of Interest}
The authors have no conflicts to disclose.

\section*{Data availability}
The data that support the findings of this study are available from the corresponding author upon reasonable request.

\bibliography{tanimura_publist,HT24,TT23}

\end{document}